\documentstyle[11pt,epsfig]{article}

\begin{document}

\setcounter{page}{0}
\newcommand{\al}{\alpha}
\newcommand{\s}{\sigma}
\renewcommand{\L}{\Lambda}
\renewcommand{\b}{\beta}
\renewcommand{\c}{\chi}
\renewcommand{\d}{\delta}
\newcommand{\D}{\Delta}
\newcommand{\wt}{\widetilde}
\renewcommand{\thefootnote}{\fnsymbol{footnote}}


\def\beq{\begin{equation}}
\def\eeq{\end{equation}}
\def\mt{m_t}
\def\GeV{\rm GeV}
\def\Im{\rm Im}
\def\ct{\tilde c_t}
\def\mutt{\left({\ct\over 2m_t}\right)}
\def\gam{\gamma}
\def\eo{\epsilon_1}
\def\et{\epsilon_2}
\def\lam{\lambda}
\def\sh{\hat s}
\def\rs{\sqrt{\hat s}}
\def\sn{\sin\Theta}
\def\cs{\cos\Theta}
\def\ssin{\sin^2\Theta}
\def\ccos{\cos^2\Theta}
\def\bh{\hat\beta}


\newcommand{\CL}{{\cal L}}
\newcommand{\CM}{{\cal M}}


\newcommand{\pl}[3]{, Phys.\ Lett.\ {{\bf #1}} {(#2)} {#3}}
\newcommand{\np}[3]{, Nucl.\ Phys.\ {{\bf #1}} {(#2)} {#3}}
\newcommand{\pr}[3]{, Phys.\ Rev.\ {{\bf #1}} {(#2)} {#3}}
\newcommand{\prl}[3]{, Phys.\ Rev.\ Lett.\ {{\bf #1}} {(#2)} {#3}}
\newcommand{\ijmp}[3]{, Int.\ J.\ Mod.\ Phys.\ {{\bf #1}} {(#2)} {#3}}
\newcommand{\mpl}[3]{, Mod.\ Phys.\ Lett.\ {{\bf #1}} {(#2)} {#3}}
\newcommand{\zp}[3]{, Z.\ Phys.\ {{\bf #1}} {(#2)} {#3}}
\newcommand{\ap}[3]{, Ann.\ Phys.\ {{\bf #1}} {(#2)} {#3}}
\newcommand{\rmp}[3]{, Rev.\ Mod.\ Phys.\ {{\bf #1}} {(#2)} {#3}}

\newpage
\setcounter{page}{0}

\begin{titlepage}
\begin{flushright}
\hfill{YUMS 97-16}\\
\hfill{SNUTP 97-081}\\
\hfill{June 1997}
\end{flushright}
\vspace{2.0cm}

\begin{center}
{\Large\bf $T$--odd Gluon--Top-Quark Effective Couplings\\
           at the CERN Large Hadron Collider}\\
\hfill{}
\vskip 0.3cm
{S.Y.~Choi, C.S.~Kim and Jake Lee}\\
{\sl Department of Physics, Yonsei University, Seoul 120-749, Korea}
\end{center}

\vskip 0.7cm
\setcounter{footnote}{0}

\begin{abstract}
The $T$--odd top--quark chromoelectric dipole moment ($t$CEDM) is probed 
through top--quark--pair production via gluon fusion at the CERN Large 
Hadron Collider (LHC) by considering the possibility of having polarized 
protons. The complete analytic expressions for the tree--level helicity 
amplitudes of $gg\rightarrow t\bar{t}$ is also presented. 
For the derived analytic results we determine the 1--$\sigma$ statistical 
sensitivities to the $t$CEDM form factor for (i) typical $CP$--odd 
observables composed of lepton and anti-lepton momenta from  
$t$ and $\bar{t}$ semileptonic decays for unpolarized protons, and 
(ii) a $CP$-odd event asymmetry for polarized protons by using the 
so-called Berger-Qiu (BQ) parametrization of polarized gluon distribution 
functions. We find that at the CERN LHC, the $CP$-odd energy and angular
correlations can put a limit of 10$^{-18}$ to $10^{-17}$ $g_s{\rm cm}$ 
on the real and imaginary parts of the $t$CEDM, 
while the simple $CP$-odd event asymmetry with polarized protons could
put a very strong limit of $10^{-20}$ $g_s{\rm cm}$ on the imaginary part
of the $t$CEDM.       
\end{abstract}

\end{titlepage}

\newpage
\renewcommand{\thefootnote}{\arabic{footnote}}
\baselineskip 24pt plus 2pt minus 2pt

The top quark with a mass of $m_t=174\pm 6$ GeV has recently been observed
at Tevatron\cite{Tevatron}.  This large mass compared to the other 
light quark masses implies that the top quark may be susceptible
to new physics effects at TeV-scale, not readily observable in 
lighter quarks.

Among the properties of the heavy top quark which should be measured 
at the CERN LHC are its basic couplings to gauge bosons.
In this Letter we investigate the possibility of extracting the $CP$-odd
gluon-top-quark effective couplings through top-quark pair production
by gluon fusion at the planned CERN LHC.

Production of $t\bar{t}$ by gluon fusion followed by $t$ and $\bar{t}$ 
semileptonic decays has been studied to extract the real part of the 
$T$-odd $t$CEDM form factor with use of optimal observables\cite{atwd}.
We extend this work\cite{atwd} by considering the possibility 
of having a nonzero imaginary part of the $t$CEDM and 
extract the imaginary as well as real parts of the $t$CEDM through 
a few typical $CP$-odd lepton and antilepton correlations\cite{bern} of the
$t$ and $\bar{t}$ semileptonic decays. 
Gunion, Yuan and Grzadkowski in Ref.~\cite{guni} have demonstrated that 
$CP$ violation in the Higgs sector can be directly probed using polarized 
gluon-gluon collisions at the CERN LHC with polarized protons. 
We apply the same method in extracting the imaginary part of the $t$CEDM 
and compare its attainable limits with those obtained through 
the $CP$-odd lepton and antilepton correlations from the $t$ and $\bar{t}$ 
semileptonic decays with unpolarized protons.

The general effective Lagrangian for gluon--top--quark interaction
includes not only the SM terms of dimension-four but also the
two terms of dimension-five:
\begin{eqnarray}
\CL_M={1\over 2}g_s\left({c_t\over 2m_t}\right){\bar t}
        \s^{\mu\nu}G^a_{\mu\nu}T_a t,\qquad 
\CL_E={i\over 2}g_s\left({\ct\over 2m_t}\right){\bar t}
        \s^{\mu\nu}\gamma_5 G^a_{\mu\nu}T_a t,
\end{eqnarray}
where $\s^{\mu\nu}={i\over 2}[\gam^\mu,\gam^\nu]$, $G^a_{\mu\nu}$ is the gluon
field strength, and $T_a={1\over 2}\lam_a$ ($a=1$ to $8$).
The coefficient, $\mu_t\equiv g_s(c_t/2m_t)$, is then called the top-quark
chromomagnetic dipole moment ($t$CMDM) which, as in the case of QED, 
receives one-loop contributions in QCD so that its size is of the order 
$g_s\al_s/\pi m_t$.
This would very likely dilute any contributions to $\mu_t$ due to new physics.
On the other hand, the coefficient in the $\CL_E$, $d_t\equiv g_s({\ct/2m_t})$,
is called the $t$CEDM, which violates $T$ invariance.  
Within the Standard Model (SM), this $t$CEDM can arise only at three or 
more loops so that it is estimated to be extremely small 
($|d_t|\le 10^{-30}g_s{\rm cm}$)\cite{dono}.  The $t$CEDM, 
however, can be much larger in certain models of $CP$ violation such as 
the models with $CP$-nonconservation through Higgs boson exchange\cite{wein},
in which the form factor may be about $10^{-20}g_s {\rm cm}$\cite{sony}.
In this light, a nonvanishing $t$CEDM should be a strong indication of new 
physics. Therefore, we will exclusively focus on the
measurement of the $t$CEDM.

Generally the $t$CEDM (and likewise the $t$CMDM) is a function 
of momentum transfer. In most model calculations, the real part of the 
$t$CEDM is nonvanishing and constant to a good approximation\cite{sony},
while its imaginary part can be nonvanishing only when the momentum transfer
is larger than the $t\bar{t}$ production threshold. 
The productions of $t\bar{t}$ pairs in proton-proton collisions 
at the CERN LHC are predominantly through the gluon-gluon fusion diagrams 
shown in Figure~1(a)-1(c). 
When we include the $t$CEDM interaction, we must include 
the Feynman diagram of Figure~1(d) in order to preserve QCD gauge invariance. 
The imaginary part of the $t$CEDM form factor in Figure~1(a) and (b)
is vanishing due to the zero momentum transfers of the on-shell gluons, 
but it might be nonvanishing in Figure~1(c) and (d) in which
the momentum transfers are larger than the $t\bar{t}$ production threshold. 
With this feature in mind, we assume for simplicity that the real and 
imaginary parts of the $t$CEDM are constant and nonvanishing. 
Certainly, the momentum-transfer dependence of the $t$CEDM should be taken 
into account in more realistic model calculations and their comparisons
with experimental data. This momentum-transfer dependence will be elaborated 
on in our future work.

The $t$CEDM term modifies the $gtt$ vertex in momentum space as
\beq
\Gamma^\mu(k,p,p')=-ig_sT_a\left[\gam^\mu+\mutt\s^{\mu\nu}\gam_5 k_\nu\right],
\eeq
where $k=p-p'$ is the gluon four-momentum,  and $p$ and $p'$ are
the four-momenta of incoming and outgoing top quarks, respectively.
In addition, QCD gauge invariance yields a dimension-five $ggt\bar t$ 
contact term, whose expression is given by
\begin{eqnarray}
\Gamma^{\mu\nu}=g_s^2 f^{abc}T_c\mutt \s^{\mu\nu}\gam_5,
\end{eqnarray}
where $f^{abc}$ is the SU(3)$_C$ group structure constants.  
This additional term describes the Feynman diagram of Figure~1(d).

%
\begin{figure}[tb]
\vspace*{-5cm}
\centerline{\epsfig{figure=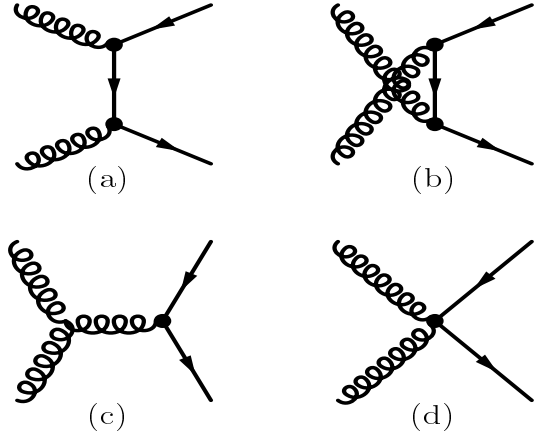,height=18cm,width=17cm,angle=0}}
\vspace*{-7cm}
\caption{\em Tree-level Feynman diagrams for $gg\rightarrow t\bar{t}$.
         The diagram of (d) is needed to preserve gauge invariance.
         The black circles denote the verticies modified by the $t$CEDM.}
\end{figure}

In evaluating the explicit form of helicity amplitudes we employ  
a set of covariant gluon polarization vectors $(\lam=\pm)$: 
\beq
\eo(\lam)=\et^*(\lam)=-{\lam \over \sqrt{2}}(n_1 +i\lam n_2),
\eeq
where $n_1$ and $n_2$ are defined as
\begin{eqnarray}
n_1^\mu={N\over 2}\left[(p_1-p_2)^\mu +{p_1\cdot(k_1-k_2)\over(k_1\cdot k_2)}
          (k_1-k_2)^\mu\right],\qquad 
n_2^\mu=N{\epsilon^{\mu\al\b\nu}p_{1\al}k_{1\b}k_{2\nu}\over(k_1\cdot k_2)},
\end{eqnarray}
with the normalization factor
\begin{eqnarray} 
N=\left[2\frac{(k_1\cdot p_1)(k_1\cdot p_2)}{(k_1\cdot k_2)}
 -m^2_t\right]^{-1/2}. 
\end{eqnarray}
These gluon polarization vectors greatly facilitate our analytic calculations
in an arbitrary reference frame due to the useful relations ($i,j=1,2$):
\begin{eqnarray}
&& n_i\cdot n_j=-\delta_{ij},\qquad k_i\cdot n_j= 0, \qquad
   p_1\cdot n_1=-p_2\cdot n_1 =-{1\over N},\nonumber\\ 
&& p_i\cdot n_2=0,\qquad\hskip 0.6cm 
   k_i\cdot \epsilon_j=0,\qquad 
   \eo(\lam_1)\cdot\et(\lam_2)=-\delta_{\lam_1\lam_2}.
\end{eqnarray}

It is now straightforward to calculate the tree-level helicity 
amplitudes for $gg\rightarrow t\bar{t}$. Noting $[T_a,T_b]=if^{abc}T_c$, 
we decompose the helicity amplitudes into two parts -  
a symmetric part and an antisymmetric part with respect to color indices,
and present their analytic expressions in the gluon-gluon c.m.~frame as 
\begin{eqnarray}
\CM_{\lam_1,\lam_2;\s,\bar\s}={2\pi\alpha_s\over 1-\bh^2\ccos}
\bigg(\{T_a,T_b\}S_{\lam_1,\lam_2;\s,\bar\s}
    +[T_a,T_b]A_{\lam_1,\lam_2;\s,\bar\s}\bigg),
\end{eqnarray}
where $\Theta$ is the scattering angle between an incoming gluon and 
a top quark, 
and $\lam_1,\lam_2$ and $\s/2,\bar\s/2$ are the helicities of two gluons,  
$t$ and $\bar t$, respectively. For convenience we introduce a $t$CEDM 
parameter $\tilde{d}_t=\tilde{c}_t/2m_t$, which is related with $d_t$ by
$d_t=g_s\tilde{d}_t$, and expand the symmetric
and asymmetric parts of the helicity amplitudes with respect to $\tilde{d}_t$
as 
\begin{eqnarray}
&& S_{\lam_1,\lam_2;\s,\bar\s}=S^0_{\lam_1,\lam_2;\s,\bar\s} 
            +(i\tilde{d}_t)S^1_{\lam_1,\lam_2;\s,\bar\s}
            +(i\tilde{d}_t)^2S^2_{\lam_1,\lam_2;\s,\bar\s},\nonumber\\
&& A_{\lam_1,\lam_2;\s,\bar\s}=A^0_{\lam_1,\lam_2;\s,\bar\s} 
            +(i\tilde{d}_t)A^1_{\lam_1,\lam_2;\s,\bar\s}
            +(i\tilde{d}_t)^2A^2_{\lam_1,\lam_2;\s,\bar\s}.
\end{eqnarray}

The SM conributions, $S^0_{\lam_1,\lam_2;\s,\bar\s}$ and 
$A^0_{\lam_1,\lam_2;\s,\bar\s}$, to the helicity amplitudes of
$gg\rightarrow t\bar{t}$ are easily evaluated and they are given by
\begin{eqnarray}
&&S^0_{\lam,\lam;\s,\s}=-{4\mt\over \rs}(\lam+\s\bh),\nonumber\\
&&S^0_{\lam,\lam;\s,-\s}=0,\nonumber\\
&&S^0_{\lam,-\lam;\s,\s}={4\mt\over\rs}\s\bh\ssin,\nonumber\\
&&S^0_{\lam,-\lam;\s,-\s}=2\bh(\lam\s+\cs)\sn,
\label{smcon_sym}
\end{eqnarray}
and
\begin{eqnarray}
A^0_{\lam_1,\lam_2;\s,\bar\s}=\bh\cs S^0_{\lam_1,\lam_2;\s,\bar\s}.
\label{smcon_asy}
\end{eqnarray}
Note that the asymmetric parts of the helicity amplitudes have the
same structure as the symmetric parts except for the helicity-independent
kinematic factor $\hat{\beta}\cos\Theta$. This remarkable factorization
property is due to a general theorem\cite{goeb}
that in any gauge theory a factorization of the internal-symmetry-(charge-) 
index dependence and the polarization (spin) dependence into separate factors 
holds for any tree-level four-particle amplitude when one or more of the 
four particles are gauge bosons. This factorization property (\ref{smcon_asy})
has been checked by an explicit calculation of both the symmetric and 
antisymmetric parts. 

The explicit form of the terms linear in $\tilde{d_t}$ are given by 
\begin{eqnarray}
&&S^1_{\lam,\lam;\s,\s}=2\rs\left[{8\mt^2\over\sh} 
                       +\bh(\bh-\lam\s)\ssin\right],\nonumber\\
&&S^1_{\lam,\lam;\s,-\s}=-4\mt\lam\bh\sn\cs,\nonumber\\
&&S^1_{\lam,-\lam;\s,\s}=2\rs\bh^2\ssin,\nonumber\\
&&S^1_{\lam,-\lam;\s,-\s}=0;\\
&&A^1_{\lam,\lam;\s,\s}={8\mt^2\over\rs}(\lam\s+\bh)\cs,\nonumber\\
&&A^1_{\lam,\lam;\s,-\s}=-4\mt\lam\sn,\nonumber\\
&&A^1_{\lam,-\lam;\s,\s}=2\rs\bh^3\cs\ssin,\nonumber\\
&&A^1_{\lam,-\lam;\s,-\s}=0,
\end{eqnarray}
and that of the terms quadratic in $\tilde{d_t}$ by 
\begin{eqnarray}
&&S^2_{\lam,\lam;\s,\s}=-2\mt\rs\lam\left[{4\mt^2\over \sh}
                       +\bh(\bh-\lam\s)\ssin\right],\nonumber\\
&&S^2_{\lam,\lam;\s,-\s}=4\mt^2\bh\sn\cs,\nonumber\\
&&S^2_{\lam,-\lam;\s,\s}=-2\mt\rs\s\bh\ssin,\nonumber\\
&&S^2_{\lam,-\lam;\s,-\s}=-\s\bh\sn\left[{4\mt^2\over\sh}\cs
                         +\lam\s(1-\bh^2\ccos)\right];\\
&&A^2_{\lam,\lam;\s,\s}=-4\s\mt^2\cs({2\mt\over\rs}),\nonumber\\
&&A^2_{\lam,\lam;\s,-\s}=4\mt^2\sn,\nonumber\\
&&A^2_{\lam,-\lam;\s,\s}=-2\s\mt\rs\bh^2\cs\ssin,\nonumber\\
&&A^2_{\lam,-\lam;\s,-\s}=-\s\bh^2\sin^3\Theta,
\label{quad_asy}
\end{eqnarray}
where $\lam,\s=\pm$, $\sqrt{\sh}$ is the gluon-gluon c.m. energy,
and $\bh=\sqrt{1-{4\mt^2\over\sh}}$. We note in passing that the
tree-level factorization, which is valid for the SM amplitudes,
is spoiled in the full amplitudes by the $t$CEDM.

$CP$ invariance requires the helicity amplitudes to satisfy the relations:
\begin{eqnarray}
\CM_{\lam_1,\lam_2;\s,\bar\s}=-(-1)^{(\lam_1-\lam_2)
              +\frac{1}{2}(\s-\bar\s)}\CM_{-\lam_2,-\lam_1;-\bar\s,-\s}.
\end{eqnarray}
From Eqs.~(\ref{smcon_sym})-(\ref{quad_asy}), we can easily check that 
the terms linear in $\tilde{d}_t$ violate $CP$, while the SM terms
and the terms quadratic in $\tilde{d}_t$ preserve $CP$. 

In the present work, we investigate the $CP$-violating effects from the 
$t$CEDM in two distinct experimental situations at the CERN LHC - 
unpolarized and polarized proton-proton collisions.

In the unpolarized case, the initial gluon-gluon configurations
with unpolarized gluons are invariant under $CP$ so that  
a detection of the $t$CEDM requires information on the top
quark polarization.  Fortunately, the left-handed nature of 
the weak decay and the very short life time\cite{bigi} of the top quark 
due to its large mass allow us to determine the top quark polarization 
quite easily. We employ the semileptonic top quark decays 
$t\rightarrow bW^+\rightarrow bl^+\nu_l$ ($l=e,\mu$), which can be easily 
identified from the formidable hadronic backgrounds in the energetic 
proton-proton collisions and which provides the most efficient handle for
top-quark polarization. The resolving power of the $t$ and $\bar{t}$
polarizations in the $t$ and $\bar{t}$ semileptonic decays in their
rest frames is determined by the decay density matrices, which are given
in the $t$ and $\bar{t}$ helicity bases by
\begin{eqnarray}
D^{\bar{l}}_t=\frac{1}{2}
          \left(
          \begin{array}{cc}
          1+\cos\theta & \sin\theta{\rm e}^{i\phi}\\
          \sin\theta{\rm e}^{-i\phi} & 1-\cos\theta
          \end{array}
          \right),\qquad 
{\bar D}^l_t=\frac{1}{2}
          \left(
          \begin{array}{cc}
          1+\cos{\bar\theta} & \sin{\bar\theta}{\rm e}^{i{\bar\phi}}\\
          \sin{\bar\theta}{\rm e}^{-i{\bar\phi}} & 1-\cos{\bar\theta}
          \end{array}
          \right),
\end{eqnarray}
respectively. Here, $\theta$(${\bar\theta}$) and $\phi$(${\bar\phi}$) 
are the polar and azimuthal angles of $l^+$($l^-$) from the $t$(${\bar t}$) 
semileptonic decay with respect to the polarization vector of positively 
polarized $t$(${\bar t}$), respectively. These decay density matrices enter
the five-fold differential cross section of the sequential process 
$gg\rightarrow t{\bar t}\rightarrow (\bar{l}\nu_l b) (l\bar{\nu}_l\bar{b})$ 
\begin{eqnarray}
{\rm d}\hat\sigma = \frac{\bh}{8\pi\sh}B_l B_{\bar l}
           {\bar\Sigma} (\Theta;\theta,\phi;\bar{\theta},\bar{\phi})
           {\rm d}\cs\frac{{\rm dcos}\theta {\rm d}\phi}{4\pi}
           \frac{{\rm dcos}{\bar\theta}{\rm d}{\bar\phi}}{4\pi},
\end{eqnarray}
through a multiplicative combination ${\bar\Sigma}$ of the production 
amplitudes and the decay density matrices as
\begin{eqnarray}
{\bar\Sigma}=\frac{1}{256}
             \sum_{\sigma,{\bar\sigma}=\pm}
             \sum_{\sigma^\prime,{\bar\sigma}^\prime=\pm}
             \sum_{\rm color} \sum_{\lambda_1,\lambda_2=\pm}
             {\cal M}_{\lambda_1,\lambda_2;\sigma,{\bar\sigma}}
             {\cal M}^*_{\lambda_1,\lambda_2;\sigma^\prime,{\bar\sigma}^\prime}
             {\bar D}^l_{\sigma\sigma^\prime} 
             D^{\bar l}_{{\bar\sigma}{\bar\sigma}^\prime},
\end{eqnarray}
where $B_{\bar l}$ and $B_l$ are the branching ratios of the $t$ and $\bar{t}$
semileptonic decays with $l=e,\mu$. This angular dependence, $\bar{\Sigma}$, 
has sixteen independent lepton and antilepton 
angular correlations, which in  principle allow us to fully reconstruct 
the information on the top-quark pair production by gluon fusion.  

In the laboratory frame, however, the gluon-gluon c.m.~frame is not fixed 
and there are two unidentifiable neutrinos so that 
the top and antitop momenta are not fully reconstructible. 
These experimental obstacles force us to employ $CP$-odd energy 
and angular correlations composed of lepton and antilepton momenta directly
measurable in the laboratory frame.  Generally, those $CP$-odd observables 
can be classified into two categories - even or odd under naive time reversal 
$\tilde{T}$, which changes the sign of the time variable, 
but does not interchange the initial and final states. 

Following the classification of Ref.~\cite{bern}, we may employ two $CP$-odd 
and $CP\tilde{T}$-even correlations as 
\begin{eqnarray}
A_1={\hat p}_g\cdot({\vec p}_{\bar l}\times {\vec p}_l),\qquad
T_{33}=2({\vec p}_{\bar l}
      -{\vec p}_l)_3({\vec p}_{\bar l}\times{\vec p}_l)_3,
\end{eqnarray}
and three $CP$-odd and $CP\tilde{T}$-odd correlations as 
\begin{eqnarray}
&& A_E=E_{\bar l}-E_l,\qquad 
   A_2={\hat p}_g\cdot({\vec p}_{\bar l}+ {\vec p}_l),\nonumber\\
&& Q^l_{33}=2({\vec p}_{\bar l}+{\vec p}_l)_3({\vec p}_{\bar l}-{\vec p}_l)_3
      -\frac{2}{3}({\vec p}^2_{\bar l}-{\vec p}^2_l).
\end{eqnarray}
However, the vector correlations $A_1$ and $A_2$ should vanish 
due to Bose symmetry of the initial $gg$ system so that these vector
correlations are not useful in detecting the $t$CEDM. So, we have to 
use $T_{33}$ 
as a $CP\tilde{T}$-even correlation proportional to the real part
of the $t$CEDM, and $A_E$ and $Q_{33}$ as $CP\tilde{T}$-odd correlations
proportional to the imaginary part of the $t$CEDM.

The quantity which should be actually evaluated is the expectation value
of a given $CP$-odd correlation $O_X$ over the $pp\rightarrow t\bar{t}$ 
production rates, whose expression is given by 
\beq
\langle O_X\rangle=\frac{\int O_Xg(x_1)g(x_2)d{\hat\sigma}dx_1dx_2}{
            \int g(x_1)g(x_2)d{\hat\sigma}dx_1dx_2}
\eeq
where $g(x)$ is the gluon distribution function, and 
$x_1$ and $x_2$ are momentum fractions of two gluons.
In order to determine its statistical significance,  
$\langle O_X\rangle$ should be compared with the statistical
fluctuation $\langle O^2_X\rangle $. When only statistical uncertainties
are taken into account, an observation different from the SM expectations 
at 1-$\sigma$ level requires  
\begin{eqnarray}
\langle O_X\rangle \geq \sqrt{\frac{\langle O^2_X\rangle }{N_{t\bar t}}},
\end{eqnarray}
for the total number of events $N_{t\bar t}=\epsilon B_l B_{\bar l}\CL_{pp}
\sigma(pp(gg)\rightarrow t{\bar t})$ with the acceptance efficiency 
$\epsilon$ and the $pp$ integrated luminosity $\CL_{pp}$.

We assume for the values of experimental parameters in our numerical 
analysis
\begin{eqnarray}
&& \epsilon=10\%,\qquad \hskip 0.6cm 
   B_l=B_{\bar{l}}=21\% \ \ {\rm for}\ \ {l=e,\mu},\nonumber\\
&& \sqrt{s}=14\ \ {\rm TeV},\ \  
   \CL_{pp}=10\ \ {\rm fb}^{-1},\ \
   m_t=175\ \ {\rm GeV},
\label{exp_para}
\end{eqnarray}
and for an unpolarized gluon distribution function employ the GRVHO 
parametrization\cite{grv}, which includes higher order corrections.

\begin{table}[htb]
\centering
\caption{\em Attainable 1-$\sigma$ limits on the real and imaginary 
             parts of the $t$CEDM, $d_t$, through the $CP$-odd 
             correlations, $T_{33}$, $A_E$ and $Q_{33}$ with the parameter 
             set (\ref{exp_para}).}
\vskip 0.5cm
\begin{tabular}{c|c}
\hline\hline
     observable & Attainable 1-$\sigma$ limits \\ \hline
       $T_{33}$ & $|Re(d_t)|=0.899\times 10^{-17} g_s{\rm cm}$\\
       $A_E$    & $|Im(d_t)|=0.858\times 10^{-18} g_s{\rm cm}$\\
       $Q_{33}$ & $|Im(d_t)|=0.205\times 10^{-17} g_s{\rm cm}$\\ 
        \hline\hline
\end{tabular}
\end{table}

Table~1 list the attainable 1-$\sigma$ limits on the real and imaginary parts 
of the $t$CEDM, $d_t$, through the $CP$-odd correlations, $T_{33}$, $A_E$ and 
$Q_{33}$ with the parameter set (\ref{exp_para}). Quantitatively, $T_{33}$ 
and $Q_{33}$ enable us to probe the real and imaginary parts of $t$CEDM 
up to the order of $10^{-17}g_s {\rm cm}$, respectively, and $A_E$ allows us 
to probe the imaginary part of the $t$CEDM down to $10^{-18}g_s {\rm cm}$. 
The numerical results for the real parts are found to be consistent with those 
in Ref.~\cite{atwd}.

Let us now move to the other situation where proton beams are polarized.
If gluons in a positively polarized proton are polarized, this polarization
transmission can be utilized to have initial $CP$-odd gluon-gluon 
configurations, which allow us to probe $CP$-violating effects in gluon 
fusion without full reconstruction of the final states. 
More concretely,  $CP$ violation is directly probed by the difference 
between its production rates through 
gluon-gluon fusion processes for colliding proton beams of opposite 
polarizations. An easily-constructed $CP$-odd asymmetry is 
the rate asymmetry $A\equiv[\s_+-\s_-]/[\s_++\s_-]$, which has been used 
as a probe of $CP$ violation in the Higgs sector\cite{guni}. 
Here, $\s_\pm$ in the process $gg\rightarrow t\bar{t}$, is the 
cross section for $t\bar t$ production in collision of an unpolarized 
proton to a proton of helicity $\pm$.

Before folding the gluon distribution functions with the cross section of the
subprocess $gg\rightarrow t\bar{t}$, we calculate the square of the matrix 
elements by summing over the final $t$ and $\bar{t}$ polarizations and,
assuming that $\tilde{c}_t$ is small and keeping the terms up to linear
in $\tilde{c}_t$, we then obtain 
\begin{eqnarray}
|\CM_{\lam_1,\lam_2}|^2
         =\frac{8\pi^2\alpha_s^2}{(1-\bh^2\ccos)^2}T_{\lam_1\lam_2}
\end{eqnarray}
where
\begin{eqnarray}
&& T_{\pm\pm}=\frac{1}{16}\left[(1+\bh^2)\pm4Im(\ct)\right]
             (\frac{7}{3}+3\bh^2\ccos),\nonumber\\
&& T_{\pm\mp}=\frac{1}{2}\bh^2\ssin(2-\bh^2+\bh^2\ccos)
             (\frac{7}{3}+3\bh^2\ccos).
\end{eqnarray}
With the above expressions we can see that 
the difference and sum of production cross sections of opposite proton
helicities are given by
\begin{eqnarray}
d\s_+-d\s_- \sim g_1\Delta g_2(T_{++}-T_{--})
            \sim g_1\Delta g_2Im(\tilde{c}_t), \qquad 
d\s_++d\s_- \sim g_1 g_2 \sum_{\lam_1,\lam_2}T_{\lam_1,\lam_2},
\end{eqnarray}
where $g_1(g_2)$ is the gluon distribution function for the unpolarized 
(polarized) proton, $\Delta g_2=g_{2+}-g_{2-}$,  
$g_{1,2}=g_{1,2+}+g_{1,2-}$, and the $\pm$ subscripts of $g_{1,2}$ and
$T_{\lam_1\lam_2}$ indicate gluons with $\pm$ helicity. So, the rate
asymmetry $A$ contains the information only on the imaginary part
of the $t$CEDM. 

Crucial for our numerical analysis is the degree of polarization that
can be achieved for gluons at the CERN LHC. The amount of gluon 
polarization in a positively polarized proton beam, defined by the 
structure function difference $\Delta g(x)=g_+(x)-g_-(x)$, is not 
currently known precisely. The relative behavior of 
$\Delta g(x)$ compared to the unpolarized gluon distribution $g(x)$ 
is theoretically constrained in the $x\rightarrow 1$
and $x\rightarrow 0$ limits: $\Delta g(x)/g(x)\rightarrow 1$ for 
$x\rightarrow 1$ and $\Delta g(x)/g(x)\propto x$ for $x\rightarrow 0$.
Several simple models which satisfy these constraints suggest\cite{brod} that a
significant amount of the proton's spin could be carried by the gluons.
The European Muon Collaboration (EMC)\cite{emc} data on the polarized 
structure function $g^p_1(x)$ is also most easily interpreted if this is the 
case\cite{ross}. Since the purpose of our work does not aim at 
the properties of the polarized gluon distribution function, we simply
employ in our analysis the BQ parametrization\cite{berg} amomg 
various parametrizations on $\Delta g$\cite{berg,chen}. 
Furthermore, we do not consider the scale 
evolution of $\Delta g$, but simply use the distribution at $Q^2=100$ 
$\GeV^2$. Analytically, the BQ parametrization is defined to satisfy
\begin{eqnarray} 
\Delta g(x)=\left\{\begin{array}{ll}
             g(x)       & (x > x_c)\\
            (x/x_c)g(x) & (x < x_c)
                   \end{array}\right.  ,
\end{eqnarray}
where $x_c\sim 0.2$ yields a value of $\Delta g\sim 2.5$ at $Q^2=10$ $\GeV^2$.
For the unpolarized gluon distribution, we use the GRVHO
parametrization as for the previous study.

To obtain a numerical indication of the observability of the asymmetry $A$, 
we include all possible $t$ decay modes so that the net branching ratio is
assumed to be unity. The statistical significance of the asymmetry can then be computed as 
\begin{eqnarray}
N_{SD}=\frac{N_+-N_-}{\sqrt{N_++N_-}} \equiv
       Im(\ct)\frac{\Delta\hat{N}}{\sqrt{N}},
\end{eqnarray}
where $N_+(N_-)$ is the number of $t\bar t$ events predicted for 
positive(negative) proton, $N=N_++N_-$, and $\Delta\hat{N}$ is defined as 
$N_+-N_-=Im(\ct)\Delta\hat{N}$. For a realistic detection efficiency 
$\epsilon$, we have only to rescale the number of events by this parameter, 
$N\rightarrow \epsilon N$. Taking $N_{SD}=1$, we obtain the 1-$\sigma$
attainable limits on the imaginary part of the $t$CEDM, $Im(d_t)$, as 
\begin{eqnarray}
|Im(d_t)|=\frac{g_s}{2m_t}\sqrt{\frac{1}{N}}\frac{N}{\Delta \hat{N}}.
\end{eqnarray}
\begin{table}[htb]
\centering
\caption{\em The number of $t\bar{t}$ events $N$, the ratio $\Delta\hat{N}/N$,
and the attainable 1-$\sigma$ limits $|\Im(d_t)|$, for $p_{_T}$-cuts with
$\sqrt{s}=14$ ${\rm TeV}$, $m_t = 175$ ${\rm GeV}$ and 
$\CL=10$ ${\rm fb}^{-1}$.}

\vskip 0.5cm 
\begin{tabular}{c|ccc}
\hline\hline
 $p_{_T}$-cuts(GeV) & $N(\times 10^6)$ & $\Delta\hat{N}/N$ 
                    & $|Im(d_t)| (\times 10^{-20}g_s{\rm cm})$\\ \hline
        0   &  2.62 & 1.44 &  0.766 \\
        20  &  2.55 & 1.42 &  0.788 \\
        40  &  2.36 & 1.37 &  0.847 \\
        60  &  2.08 & 1.30 &  0.951 \\
        80  &  1.74 & 1.22 &  1.107 \\
        100 &  1.41 & 1.14 &  1.313 \\ \hline\hline
\end{tabular}
\end{table}

Experimentally, a sizable transverse-momentum cut may be needed
to reduce the formidable hadronic backgrounds and so we take into account 
the $p_{_T}$-cut dependence in determining
the number of $t\bar t$ events $N$, the ratio $\Delta\hat{N}/N$ 
and the 1-$\sigma$ limits of $|Im(d_t)|$ for an integrated 
luminosity of $10$ ${\rm fb}^{-1}$. There is, however, no drastic 
$p_{_T}$-cut dependence as can be clearly seen in Table~2. 
A large $p_{_T}$ can be, therefore, taken to reduce large 
amount of background effects without spoiling the attainable limits on
the $t$CEDM. Remarkably, the attainable 1-$\sigma$ limits
for the imaginary part, $Im(d_t)$, are much more stringent than those
obtained through the lepton and antilepton correlation $A_E$ of $t$ and 
$\bar{t}$ decay products (See Table~1.), although the same value of 
acceptance efficiency is taken in the polarized case. 
Numerically, it is possible to put a limit on the imaginary part of 
the $t$CEDM, $Im(d_t)$, up to the order of 10$^{-20}$ $g_s{\rm cm}$ 
at 1-$\sigma$ level.

Certainly, more precise theoretical estimates and experimental determinations
of polarized gluon distribution in polarized proton 
must be performed.  If $\Delta g(x)$ is precisely known, then 
detection of $A$ is relatively straightforward. At any rate, taking into
account the theoretical constraints on the $x\rightarrow 0$ and 
$x\rightarrow 1$ limits of $\Delta g(x)$, and various model parametrizations, 
we do not regard our choice of the simple BQ parametrization as 
unlikely.  Consequently, the ability of polarizing one of
the proton beams at the CERN LHC could provide a unique opportunity for
detecting $CP$ violation in $gg\rightarrow t\bar{t}$ as well as in
the Higgs sector.

To summarize, the real and imaginary part of the $t$CEDM 
can be measured through the top-quark pair production by gluon fusion
to a precision of the order of 10$^{-18}$ $g_s{\rm cm}$
by use of the lepton and antilepton correlations of the $t$ and $\bar{t}$
semileptonic decays in unpolarized proton-proton collisions, and the imaginary
part of the $t$CEDM to a precision of the order of 10$^{-20}$ $g_s{\rm cm}$ 
by use of the asymmetry between production rates for positively versus 
negatively polarized protons, if a reasonable amount of the proton polarization 
is transmitted to the gluon distributions. This large enhancement with proton 
polarization could be a reasonable motivation for having polarized LHC beams.

\vskip 2.0cm
\centerline{\bf Acknowledgements}
\medskip

The authors would like to thank Prof. H.S.~Song and Dr. M.S.~Baek for 
careful reading and useful comments.
The work was supported in part by the KOSEF-DFG large collaboration 
project, Project No. 96-0702-01-01-2.
The work of SYC was supported in part by the Korean Science and Engineering  
Foundation (KOSEF) and Korean Federation of Science and Technology
Societies through the Brain Pool program. 
The work of CSK and JL was supported 
in part by the CTP, Seoul National University, 
in part by Yonsei University Faculty Research Fund of 1997, and
in part by the BSRI Program, Ministry of Education, Project No. BSRI-97-2425.

\end{document}